\documentclass[a4paper,aps,prd,10pt,preprintnumbers,showpacs,twocolumn,superscriptaddress,nofootinbib,amsmath,amssymb,floatfix]{revtex4-1}\def\JCAPstyle#1{}
\usepackage{enumitem}
\usepackage{graphicx}
\usepackage[utf8]{inputenc}
\usepackage[T1]{fontenc}
\usepackage{cmap}
\usepackage{subfigure}
\usepackage{color}

\DeclareMathAlphabet{\pazocal}{OMS}{zplm}{m}{n}

\begin{document}

\preprint{APS/-QNM}

\title{Critical behavior and Joule-Thomson expansion of charged AdS black holes surrounded by exotic fluid with modified Chaplygin equation of state}
\author{Meng-Yao Zhang}
\email{gs.myzhang21@gzu.edu.cn}
\affiliation{School of Mathematics and Statistics, Guizhou University, Guiyang, 550025, China}
\author{Hao Chen}
\email{haochen1249@yeah.net }
\affiliation{School of Physics and Electronic Science, Zunyi Normal University, Zunyi 563006,PR  China}
\author{Hassan  Hassanabadi}
\email{h.hasanabadi@shahroodut.ac.ir}
\affiliation{Faculty of Physics, Shahrood University of Technology, Shahrood, Iran}
\affiliation{Department of Physics, University of Hradec Kr\'{a}lov\'{e}, Rokitansk\'{e}ho 62, 500 03 Hradec Kr\'{a}lov\'{e}, Czechia}

\author{Zheng-Wen Long}
\email{zwlong@gzu.edu.cn (Corresponding author) }
\affiliation{College of Physics, Guizhou University, Guiyang, 550025, China}

\author{Hui Yang}
\email{huiyang@gzu.edu.cn (Corresponding author)}
\affiliation{School of Mathematics and Statistics, Guizhou University, Guiyang, 550025, China}

\date{\today }
\begin{abstract}
		\textbf{Abstract:} By considering the concept of a unified single fluid model, referred to as the modified Chaplygin gas (MCG), which amalgamates dark energy and dark matter. In this study, we explore the thermodynamic characteristics of charged anti-de Sitter (AdS) black holes existing in an unconventional fluid accompanied by MCG. To accomplish this objective, we derive the equations of state by regarding the charge $Q^{2}$ as a thermodynamic variable. The impact of MCG parameters on the critical thermodynamic quantities ($\psi_{c}$, $T_{c}$, $Q_{c}^{2}$)  are examined, followed by a detailed analysis of the $Q^{2}-\psi$ diagram. To provide a clearer explanation of the phase transition, we present an analysis of the Gibbs free energy. It is important to note that if the Hawking temperature exceeds the critical temperature, there is a distinct pattern observed known as swallowtail behavior. This indicates that the system undergoes a first-order phase transition from a smaller black hole to a larger one. The critical exponent of the system is found to be in complete agreement with that of the van der Waals  fluid system. Furthermore, we investigate the impact of MCG parameters and black hole charge on the Joule-Thomson (J-T) expansion in the extended phase space. The J-T coefficient is examined to pinpoint the exact region experiencing cooling or heating, the observation reveals that the presence of negative heat capacity results in the occurrence of a cooling process. The impact of MCG on the inversion curve of charged black holes exhibits a striking resemblance to that observed in most multi-dimensional black hole systems. In addition, it is worth noting that the certain parameters exert a significant influence on the ratio $\frac{T_{min}}{T_{c}}$. For the specific values of MCG parameters, the ratio is consistent with the charged AdS black hole. The parameters $\gamma$ and $\beta$ have a non-negligible effect on the isenthalpy curve.

	
\end{abstract}
	
	\maketitle
	
	\section{Introduction}
The thermodynamic characteristics of black holes have become a pivotal intersection in the domains of general relativity (GR), thermodynamics, and quantum gravity. The discovery made by Hawking regarding the thermodynamic emission of particles from black holes has significantly heightened interest in these celestial objects, leading to extensive investigations into their various thermodynamic properties \cite{cq1,cq2,cq3,cq4,cq5}. The pioneering research by Hawking and Page explored the thermodynamics of anti-de Sitter (AdS) black holes, revealing a phase transition from a stable Schwarzschild black hole \cite{cq6}. Furthermore, this phase transition can also be extended to  include charged AdS black hole, the findings indicate that this phenomenon is associated with the van der Waals fluid system \cite{cq7,cq8}.  The interpretation of the cosmological constant is crucial in exploring the thermodynamic black holes, which is regarded as the pressure existing within the black hole that precisely aligns with the Smarr formula, the black hole mass  is regarded as the  enthalpy instead of its internal energy, and the thermodynamic volume of the conjugate pressure can be obtained using laws of black hole thermodynamics \cite{cq10}.

The study of phase transitions in the extended phase space for black holes has attracted considerable attention, leading to the discovery of numerous thermodynamic phase transition phenomena and phase structures. Such as, Kubiznak and Mann studied the critical behavior of an AdS black hole , and made a precise comparison between the phase transition from small to large black holes with liquid-gas systems \cite{cq9}. Reentrant phase transition and triple point have been observed in various AdS black hole configurations \cite{R1, R2}. Subsequently, Wei et al. identified a remarkable triple points within the charged Gaussian-Bonnet black hole \cite{cq10.5} , investigated the reentrant phase transition of a single-spin black holes, and analyzed the impact of different dimensions on the critical point \cite{cq11}. The investigation of the second-order phase transition of black holes involves the introduction of the notion of number density for black hole molecules, aiming to enhance our novel comprehension of the microstructure \cite{cq12}. A novel methodology was introduced by Hendi et al. to determine the critical pressure and horizon radius of black holes, facilitating the identification of the specific phase transition occurring in black holes \cite{cq13,cq14}. Recently, there has been extensive research on the thermodynamic topological properties of various black holes based on Duan's $\phi$-mapping theory \cite{cq15,cq16,cq17,cq18,cq19,cq20,cq21,cq22,cq23,my}, which contributes to a more profound comprehension of the thermodynamic characteristics of black holes.

The cosmological constant is used to define the pressure exerted by a black hole, considerable advancements have been achieved in the examination of black hole phase transition. However, in the context of general relativity theory, the cosmological constant is regarded as a constant value associated with the AdS geometric background, the interpretation of it as a variable pressure would be conceptually flawed. On the other hand, when considering $Q$ as a thermodynamic variable, there will be similarities to van der Waals phase transitions and critical phenomena \cite{cq7,cq8}. However, this approach poses mathematical challenges and deviates from conventional physical principles. In this regard,the authors, Sheykhi et al., have developed a novel Smarr formula by substituting $Q^{2}$ with $Q$ to rectify the principle of black hole thermodynamics within an alternative phase space \cite{ch7}. The identification of the critical behavior and occurrence of a first-order phase transition in the Lifshitz expansion black hole has been achieved successfully \cite{ki1}. Similarly, the investigation of thermodynamic phase transitions of other black holes in the alternative phase space has been extensively explored \cite{ki2,ki3,ki4,ki5}.

The classical thermodynamic concept of Joule Thomson (J-T) expansion is commonly recognized as the migration of gas from a region of higher pressure to one of lower pressure at an equivalent speed during the process, there is no alteration in enthalpy. \"Ozg\"ur \"Okc\"u and Ekrem Aydiner initially applied the principles of J-T expansion to investigate the thermodynamics of black holes, they conducted a comprehensive analysis on the J-T effect exhibited by charged AdS black holes, resulting in the derivation of  inversion temperature and curve. The distinct areas associated with the cooling and warming processes were effectively identified through the utilization of the $T-P$ diagram \cite{ch11,R3}. Consequently, the J-T expansion rapidly gained widespread popularity among various categories of black holes, such as the effects of the quintessence dark energy \cite{R4}, space-time dimension \cite{R5}, angular momentum \cite{R6} and Rainbow gravity \cite{R7} on J-T expansion. We investigated the effects of the space-time dimension on the minimum inversion temperature and cooling-heating effect for charged dilatonic black hole \cite{ww2}. For more related discussion, see \cite{ww1,ww3,ww4,ww5,ww6}.

In the context of the standard cosmological model, astronomical observations have confirmed that dark energy (DE) constitutes approximately $73\%$ of the universe, while dark matter (DM) makes up around $23\%$, and baryonic matter accounts for a mere $4\%$ \cite{ki6}. The different models of dark energy have been proposed, offering both empirical and theoretical predictions to elucidate the universe's accelerated expansion, such as quintessence and quintom models, in order to explain these observations \cite{ki7,ki8}. In an endeavor to offer a more comprehensive elucidation for the accelerated expansion of the Universe, researchers have proposed a novel  mixing dark matter and dark energy. Specifically, they have suggested employing the Chaplygin gas (CG) model \cite{ki9}, such as Generalized Chaplygin gas (GCG) \cite{ki10,ki11}. The effect of Chaplygin-like fluids on black hole spacetime was first considered in reference \cite{R8}. Subsequently, such a model was extended to the modified Chaplygin gas (MCG) \cite{ki12,ki13}. The thermodynamic phase transition behavior \cite{R9} and shadows \cite{R10} of black holes in Chaplygin-like dark fluid are also discussed. It is worth mentioning that the relativistic framework has been employed to address the charged AdS black hole with MCG. The author conducted an investigation into the criticality of P-V and observed a similarity between the phase transition of small/large black holes and the liquid/gas phase transition in van der Waals \cite{ki14}. To advance our understanding of the thermodynamic properties of a charged AdS black hole enveloped by MCG, this study aims to analyze the fixed parameter associated with the cosmological constant and explore criticality through an alternative perspective using a novel equation-of-state pair featuring variable $Q^{2}$ charge. Furthermore, it will undertake an investigation into how black hole characteristics affect their influence on J-T expansion.

The arrangement of this paper is outlined below: In Section II, we provide a concise overview of the solution for the charged AdS black hole with MCG, a comprehensive analysis is conducted to examine the impact of MCG parameters on the critical behavior and the Gibbs free energy associated with black holes. In Section III, we initially analyze the impact of the black hole charge Q on both the J-T coefficient and Hawking temperature, subsequently deriving the inversion temperature and pressure for a black hole surrounded by MCG. In conjunction with visual figures, we thoroughly discuss the variation trends of MCG parameters concerning the ratio $\frac{T_{min}}{T_{c}}$  and isenthalpic curves. The latter section summarizes our findings.

	\section{Thermodynamic of a charged AdS black holes surrounded by MCG}\label{sec2}
In this particular section, we direct our attention towards an AdS black hole with a MCG structure within the context of GR, as elucidated by the subsequent action \cite{ki14}
\begin{equation}
\mathcal{I}=\frac{1}{16 \pi} \int \mathrm{d}^4 x \sqrt{-g}\left[\mathcal{R}+6 \ell^{-2}-\frac{1}{4} F_{\mu \nu} F^{\mu \nu}\right]+\mathcal{I}_M,
\label{1}
\end{equation}
the Ricci scalar $\mathcal{R}$, the determinant of the metric tensor $g_{\mu\nu}$ denoted as $g=\det (g_{\mu\nu})$, the AdS length represented by $\ell$, and the field strength of the electromagnetic field given by $F_{\mu\nu}=\partial_\mu A_\nu-\partial_\nu A_\mu$ with $A_\mu$ being the gauge potential are all involved in this context.  $\mathcal{I}_M$ is the matter contribution arising from the MCG background. MCG
is an extension of the Chaplygin gas, which is widely regarded as a perfect exotic fluid. In this study, we consider the following form of MCG.
\begin{equation}
p=A \rho-\frac{B }{ \rho^\beta},
\end{equation}
where the values of $A$ and $\beta$ must satisfy the condition of being non-negative, $A$ and $B$ are considered as constants in the adiabatic processes \cite{ch1}.
If specific values are assigned to the parameters, the MCG will transition into alternative models of DM and DE. On the one hand, the conditions $\beta >1$ and $A = 0$ can be transformed into the concept of GCG: $p=-\frac{B }{ \rho^{\beta}}$. On the other hand, if the parameters satisfy the condition $\beta =1$ and $A = 0$, the MCG will degenerate into a simple pure CG: $p=-\frac{B }{ \rho}$, which accurately  characteristic of the aerodynamic process generates lifting forces on the wing of an aircraft.
 Varying the action $(\ref{1})$ leads to the following field equations:
 \begin{align}
    G_{\mu\nu}-\frac{3}{\ell^2}g_{\mu\nu}&=T_{\mu\nu}^{\text{EM}}+T_{\mu\nu}^{\text{CMG}}\label{G},\\
    \partial_\mu(\sqrt{-g}F^{\mu\nu}) &=0,
    \label{2}
\end{align}
the Einstein tensor $G_{\mu\nu}$, along with $T_{\mu\nu}^{\text{CMG}}$ refers to the energy-momentum tensor associated with MCG, while $T_{\mu\nu}^{\text{EM}}$ represents the specific formulation of the energy-momentum tensor related to the electromagnetic field
\begin{equation}
T_{\mu\nu}^{\text{EM}}=2\left(F_{\mu\lambda}F_\nu^\lambda-\frac{1}{4}g_{\mu\nu}F^{\lambda\delta}F_{\lambda\delta}\right).
    \label{3}
\end{equation}
In this case, we consider a four-dimensional space-time that is static and exhibits spherical symmetry
\begin{equation}
ds^2 = -f(r) \,\mathrm{d} t^2 + \dfrac{\mathrm{d} r^2}{f(r)} +r^2 \mathrm{d}\Omega^2,
\label{2}
\end{equation}
 the metric function $f(r)$ is dependent on the variable $r$, and $d \Omega^2=d \theta^2+\sin ^2 \theta d \phi^2$. To accurately determine the expression for energy density, we provide the metric function in equation $(\ref{2})$ and consider applying conservation conditions to the energy momentum tensor, this consideration yields an intriguing result, such as \cite{ki12,ki14}
\begin{equation}
    \rho(r)=\Bigg\{\frac{1}{1+A}\Bigg(B+\bigg(\frac{\gamma}{r^3}\bigg)^{(1+A)(1+\beta)}\Bigg)\Bigg\}^{\frac{1}{1+\beta}},
    \label{22}
\end{equation}
the utilization of an integration constant, denoted as $\gamma>0$, is crucial. It should be noted that the energy density is confined to approximately $\left(\frac{B}{1+A}\right)^{\frac{1}{1+\beta}}$. As a consequence, MCG exhibits characteristics akin to a cosmological constant in regions distant from the black hole, while progressively manifesting increased gravitational density as it nears the black hole.

Now, we can derive a concise expression for the analytical solution $f(r)$
\begin{eqnarray}\nonumber
        f(r) &=& 1-\frac{2
   M}{r}+\frac{Q^2}{r^2}+\frac{r^2}{\ell^2}\\ &-&\frac{r^2}{3} \bigg(\frac{B}{A+1}\bigg)^{\frac{1}{\beta+1}} \,
   _2F_1[\alpha, \nu; \lambda; \xi],
   \label{3}
    \end{eqnarray}
    here, the variables $M$ and $Q$ denote the physical attributes of black hole mass and charge, respectively, and the hypergeometric function ${_2}F_1[\alpha, \nu; \lambda; \xi]$  can be expressed as a power series
\begin{equation}
{_2}F_1[\alpha, \nu; \lambda; \xi]=\sum_{k=0}^{\infty}\bigg[(\alpha)_k(\nu)_k/(\lambda)_k\bigg]\xi^k/k!,
\end{equation}
where $\lvert \xi\rvert<1$, the Pochhammer symbol is denoted by $(n)_k$ \cite{ch3}, the parameter sets $(\alpha, \nu, \lambda, \xi)$ can be expressed as
    \begin{align}
      \alpha&=-\frac{1}{\beta+1},
      \,\,\,\nonumber
      \nu=-\frac{1}{1+A +\beta( A+1)}, \\
     \lambda&=1+\nu, \,\,\hspace{0.5cm}
     \xi=-\frac{1}{B}\left(\frac{\gamma}{r^3}\right)^{(A+1) (\beta+1)}.\nonumber
    \end{align}
Next, our focus will be on examining the phase transition and critical behavior. In this end, by utilizing the line element presented in Eq.\eqref{3}, the mass at the horizon radius can be written as
    \begin{eqnarray}\label{Mass}\nonumber
   M=\frac{1}{6 r_+}\Bigg\{
  &-&r_+^4\,\bigg(\frac{A+1}{B}\bigg)^{\alpha}   \, _2F_1[\alpha, \nu; \lambda; \xi] \\
  &+&\frac{3}{\ell^2}r_+^4+ 3( Q^2 +r_+^2) \Bigg\}.
  \end{eqnarray}
The investigation of black hole thermodynamic phase transition is commonly conducted within the extended phase space, where cosmological constant and the corresponding conjugate quantities are defined as pressure and volume \cite{ch2.1,ch2.2}
\begin{equation}
 P=-\frac{\Lambda}{8\pi}=\frac{3}{8\pi l^{2}},
\end{equation}
here, $l$ represents the radius of AdS, while $\Lambda$ denotes a cosmological constant with negative value. To calculate the Hawking temperature, one must initially take into account the surface gravity \cite{ch4}, specifically, it can be expressed as
\begin{equation}\label{R1}
\begin{aligned}
T& =\left.\frac{1}{4 \pi} \frac{d f(r)}{d r}\right|_{r_{+}} \\ &= \frac{1}{4 \pi r_{+}}+2 P r_{+}-\frac{\left(\frac{1+A}{B}\right)^\alpha(1-\xi)^{-\alpha} r_{+}}{4 \pi} \\
& -\frac{Q^2}{4 \pi r_{+}^3}.
\end{aligned}
\end{equation}
The thermodynamic quantities of the system can be determined by considering the first law of black hole thermodynamics \cite{ch4.4}
\begin{equation}
\mathrm{d} M=T_{+} \mathrm{d} S+\sum_i \mu_i \mathrm{~d} \mathcal{N}_i,
\end{equation}
here $\mu_{i}$ denote the chemical potentials linked to the conserved charges $\mathcal{N}i$. The area of the event horizon can be written as:
\begin{equation}
\begin{split}
A &= \int \int \sqrt{g_{\theta\theta}g_{\varphi\varphi}}\,d\theta d\varphi\\
& =\int^{\pi}_{0} \int^{2\pi}_{0} r_{+}^{2}\sin\theta \,d\theta d\varphi \\
& =4 \pi r_{+}^{2}
\end{split}
\end{equation}
we can calculate the entropy of the black hole as follows:
\begin{equation}\label{Mas1}
S = \frac{A}{4} = \pi r_{+}^{2}
\end{equation}
Next, we will explore the phase transition of the system in a novel phase space. Specifically, the cosmological constant is fixed while considering the $Q^{2}$ as the thermodynamical variable. To this end, by incorporating Eq. \eqref{Mas1}  into the mass expression, we can obtain
\begin{equation}\label{R2}
\begin{aligned}
M & =\frac{\left(\frac{1+A}{B}\right)^{\alpha}}{6 \pi^{3 / 2}} S^{3 / 2} { }_2 F_1\left[\alpha, \nu, \lambda ; \frac{\pi^{3 / 2}}{B}\left(\frac{\gamma}{S^{3 / 2}}\right)^{(1+A)(1+\beta)}\right]\\
& +\frac{\pi Q^2}{2 \sqrt{\pi S}}+\frac{4 P S^{3 / 2}}{3 \sqrt{\pi}}+\frac{\sqrt{S}}{2 \sqrt{\pi}}.
\end{aligned}
\end{equation}
Based on this, we can obtain the  conjugate variables as follows
\begin{equation}
T=\left.\frac{\partial M}{\partial S}\right|_{P, Q^2}, \psi=\left.\frac{\partial M}{\partial Q^2}\right|_{P, S}=\frac{1}{2 r_{+}},
\end{equation}
\begin{equation}
V=\left.\frac{\partial M}{\partial P}\right|_{Q^2, S}=\frac{4 \pi}{3} r_{+}^3,
\end{equation}
here $T$ denotes the temperature. The aforementioned thermodynamic quantities adhere to the principles of the first law of thermodynamics within the new phase space
\begin{equation}
d M=T d S+\psi d Q^2+V d P.
\label{23}
\end{equation}
The consideration of charge $Q$ as a thermodynamic variable poses mathematical challenges and deviates from conventional physical principles \cite{cq7,cq8}. In other words, the variables $Q$ and $\phi$ ($\phi=\partial M /\partial Q$) are not mathematically independent since $\phi=Q/r_+$, which can lead to physically irrelevant quantities such as $\partial Q /\partial \phi$. This selection is conducive to the relation
\begin{equation}
\phi=\frac{Q}{r_{+}} \Rightarrow \phi Q=\frac{Q^2}{r_{+}}.
\end{equation}
 Combined with Eqs.\eqref{R1} and \eqref{R2}, when considering fixed $\Lambda$, the term $\psi dQ^{2}$ (i.e. $Q^{2}$ as a thermodynamic variable) is substituted for the term $\phi dQ$ as indicated by Eq.\eqref{23} is the best choice. We then substitute $r_{+}=1 /(2 \psi)$ into Hawking temperature, resulting in the equation of state for  $Q^2(T, \psi)$ as follows:
\begin{equation} \label{JJ3}
\begin{aligned}
 Q^2(T, \psi)&=\frac{3}{16l^2 \psi^4}-\frac{\pi T}{2 \psi^3}+\frac{1}{4 \psi^2}-\frac{\left(\frac{1+A}{B}\right)^\alpha}{16 \psi^4}\\
&\times\left(1+\frac{8^{(1+A)(1+\beta)}\left(\gamma \psi^3\right)^{(1+A)(1+\beta)}}{B}\right)^{-\alpha}.
\end{aligned}
\end{equation}
\begin{figure}
		\centering
		\includegraphics[scale = 0.5]{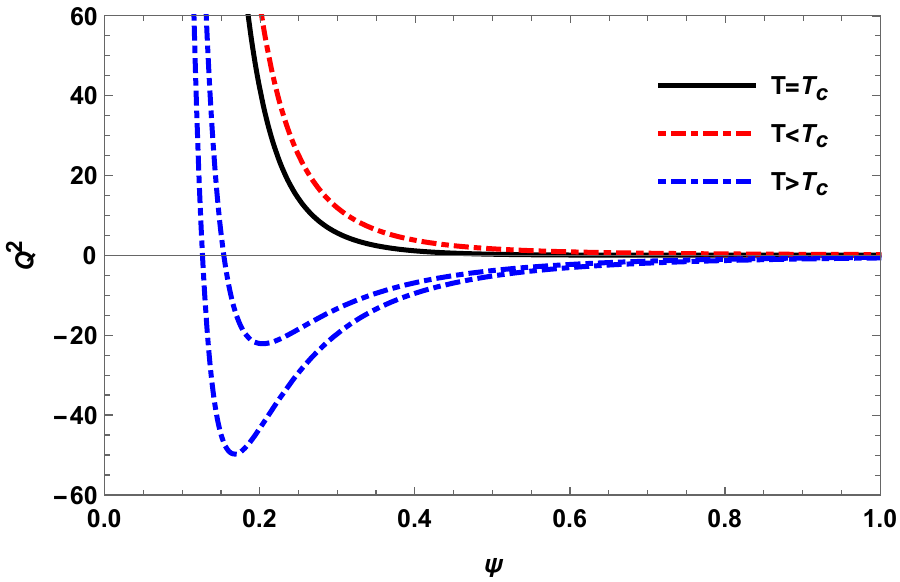} \hspace{-0.2cm}
		\includegraphics[scale = 0.5]{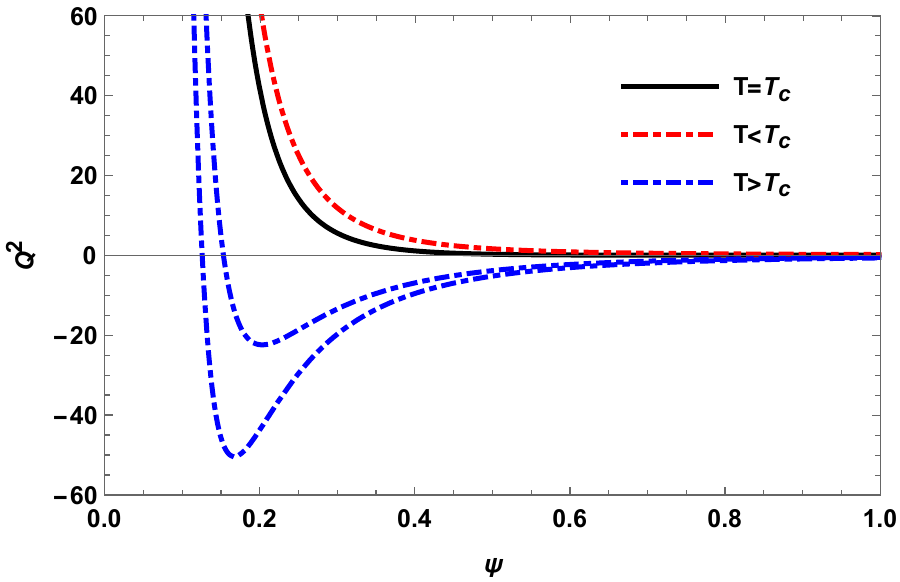} \vspace{-0.2cm}
        \includegraphics[scale = 0.5]{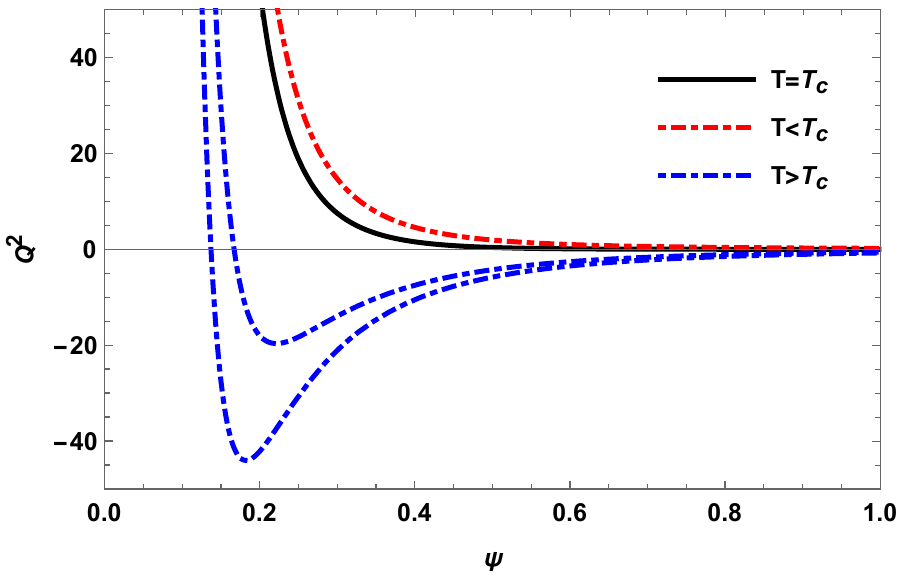} \vspace{-0.2cm}
		\caption{ $Q^{2}-\psi$ diagram.  Up panel: ($\beta=B=l=1,A=0,\gamma=0.1$). Middle panel: ($B=l=1,\beta=2, A=0,\gamma=0.1$). Down panel: ($A=B=l=1,\beta=\gamma=0.1$).}
		\label{fig:WKBL0}
	\end{figure}
The isothermal diagram of a charged black hole in the $Q^{2}-\psi$ plane is presented in FIG.1, illustrating the physical makeup acceptance region. When the Hawking temperature falls below the critical temperature $(T_{c}=0.2169; T_{c}=0.2175; T_{c}=0.2458)$, an anomalous negative $Q^{2}$ component can be observed within this region on the isothermal diagram. This phenomenon bears resemblance to that observed in a typical van der Waals fluid, where pressure may also exhibit negativity at certain values \cite{ch7,ch8}. The inherent instability of these features can be effectively mitigated through the conventional Maxwellian equal area construction
\begin{equation}
\oint \psi d Q^2=0,
\end{equation}
the presence of an critical point signifies the occurrence of phase transition, the critical point is determined by
\begin{equation}
\left.\frac{\partial Q^2}{\partial \psi}\right|_{T_c}=0,\left.\quad \frac{\partial^2 Q^2}{\partial \psi^2}\right|_{T_c}=0,
\end{equation}
the  black hole thermodynamic critical quantity ($T_c, Q_c^2$) can be expressed as
\begin{table}[]
		\caption{Critical values for $l=1$ with different
parameters}
		\centering
		\label{Table:remnant}
		\begin{tabular}{|l|l|l|l|l||l||l|}
			\hline
			$A$ & $\beta$ & $B$ & $\gamma$ & $\psi_{c}$& $T_{c}$& $Q^{2}_{c}$\\ \hline\hline
			0 & 1 & 0.1 & 0.01 & 1.14918& 0.24629& 0.02975 \\ \hline
			0 &1 & 0.1 & 0.1 & 1.40746 &	0.23993 &	0.00292 \\ \hline
			0&	1	&1&	0.01&	0.99805	&0.21229	& 0.04134\\ \hline
			0&	1	&1&	0.1&	0.94249	&0.21685&	 0.01686 \\ \hline\hline
			0	&0.1&	1	&0.01&	1.00592&	0.21167&	0.03898 \\ \hline
			0&	0.1&	0.1&	0.1	& 1.41903&	0.23800	& 0.00304 \\ \hline
			0&	0.1&	1&	0.1	& 1.11226	&0.20416	& 0.01097 \\ \hline
			0&	2	&0.1&	0.1	 & 1.42095&	0.24076	& 0.00271 \\ \hline\hline
			0.1&	0.1&	0.1	&0.01&	 1.20822&	0.25448&	 0.02656 \\ \hline
			0.1	&0.1&	0.1&	0.05&	 1.25885&	0.25086	& 0.01497 \\ \hline
			0.1	&1&	0.1&	0.05&	 1.18060&	0.24977&	 0.01579 \\ \hline
			1&	0.1	&0.1	&0.05&	 1.11785&	0.26206&	 0.01665 \\ \hline
			1&	0.1&	1&	0.05&	 1.04269&	0.23892	& 0.02490\\ \hline\hline
		\end{tabular}
	\end{table}
\begin{equation}
\begin{aligned}
T_c & =\frac{\left(\frac{1+A}{B}\right)^\alpha(1-\xi)^{-1-\alpha} \xi}{24 \pi \psi_c} \frac{\left(\frac{1+A}{B}\right)^\alpha(1-\xi)^{-1-\alpha}}{6 \pi \psi_c }\\
& -\frac{A\left(\frac{1+A}{B}\right)^\alpha(1-\xi)^{-1-\alpha} \xi}{8 \pi \psi_c}+\frac{\psi_c}{3 \pi}+\frac{1}{21^2 \pi\psi_c},
\end{aligned}
\end{equation}
\begin{equation}
\begin{aligned}
Q_c^2 & =\frac{\psi_c^2}{3}+\frac{1}{3}\left(\frac{1+A}{B}\right)^\alpha(1-\xi)^{-\alpha}\psi_c^4 -\frac{\psi_c^4}{l^2}\\
& -\frac{\left(\frac{1+A}{B}\right)^{-\alpha \beta} B(1-\xi)^{-\alpha} \xi \psi_c^4}{-1+\xi}.
\end{aligned}
\end{equation}
In this case, some numerical calculations are necessary to study the impact of parameters on their phase transitions. As shown in TABLE.I, we observe that
\begin{enumerate}[label=\roman*.]

\item For $(A=0,\beta=1)$, $T_{c}$ and $Q_{c}^{2}$ decrease with an increase in parameter $\gamma$ while $\psi_{c}$ exhibits the opposite trend. The increase of parameter $B$ leads to a decrease in both $\psi_{c}$ and $T_{c}$, while $Q_{c}^{2}$ experiences an increase. The phase transition can be more readily achieved by increasing the parameter $\gamma$, as compared to $B$.

\item For  $(A=0,\beta=2)$, the results indicate that the values of $\psi_{c}$ and $T_{c}$ exhibit an increasing trend with higher $\beta$, whereas $Q_{c}^{2}$ demonstrates an inverse relationship.

\item For $(A\neq0)$, the values of $T_{c}$ and $Q_{c}^{2}$ increase as parameter $A$ increases, while the growth trend of $\psi_{c}$ is inhibited by A. It should be noted that changing parameter $B$ significantly alters the critical thermodynamic quantity of the black hole, resulting in a reversed $CG$ contrast.
\end{enumerate}
The characterization of the phase transition can be effectively achieved through the Gibbs free energy, which is expressed as follows:
\begin{equation}
\begin{aligned}
G & =M-T S \\
& =\frac{3 Q^2}{4 r_{+}}+\frac{r_{+}}{4}-\frac{r_{+}^3}{4l^2}+\frac{1}{4}\left(\frac{1+A}{B}\right)^\alpha(1-\xi)^{-\alpha} r_{+}^3 \\
& -\frac{1}{6}\left(\frac{1+A}{B}\right)^\alpha{ }_2 F_1[\alpha, v ; \lambda ; \xi] r_{+}^3.
\end{aligned}
\end{equation}
In FIG.2, we have given the $G-Q^{2}$ diagram with critical values of thermodynamic quantities
($ (i)\beta=l=1,A=0,B=0.1,\gamma=0.01$,$\psi_{c}=1.14918, T_{c}=0.24629,Q^{2}_{c}= 0.02975$; $(ii)\beta=2,l=1,A=0,B=0.1,\gamma=0.01$,$\psi_{c}=1.12122, T_{c}=0.23908,Q^{2}_{c}= 0.03263$;$(iii) A=B=\beta=0.1,l=1,\gamma=0.01$,$\psi_{c}=1.20822, T_{c}=0.25448,Q^{2}_{c}= 0.02659$ ), note that the Hawking temperature is less than or equal to the critical temperature, the charged black hole is stable, which can also be verified from the $Q^{2}-\psi$ diagram in FIG.1. However, when the temperature of the system exceeds a certain critical point, it displays a characteristic swallowtail behavior which suggests that there is a first-order phase transition taking place.
\begin{figure}
		\centering
		\includegraphics[scale = 0.32]{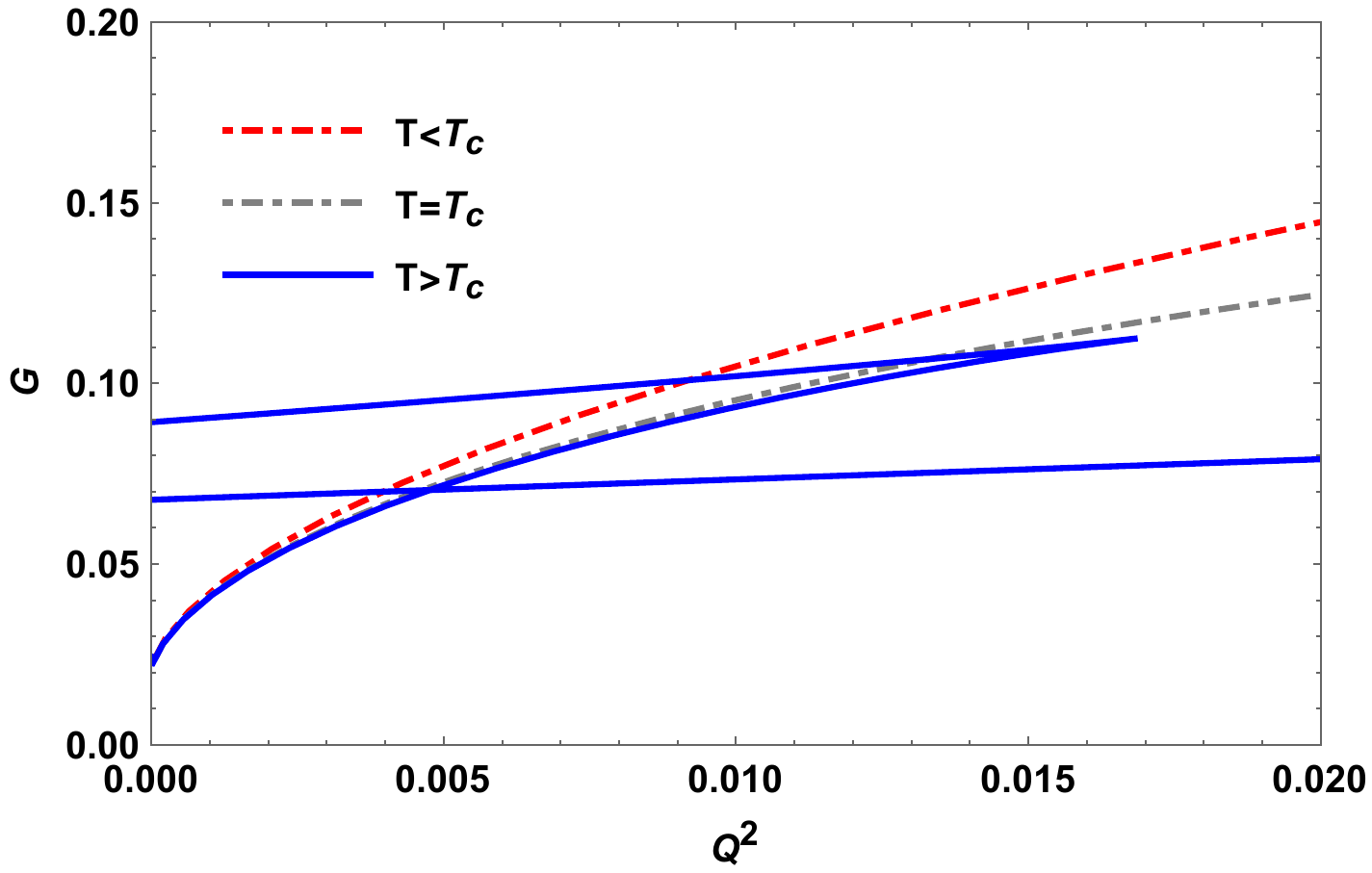} \hspace{-0.2cm}
		\includegraphics[scale = 0.32]{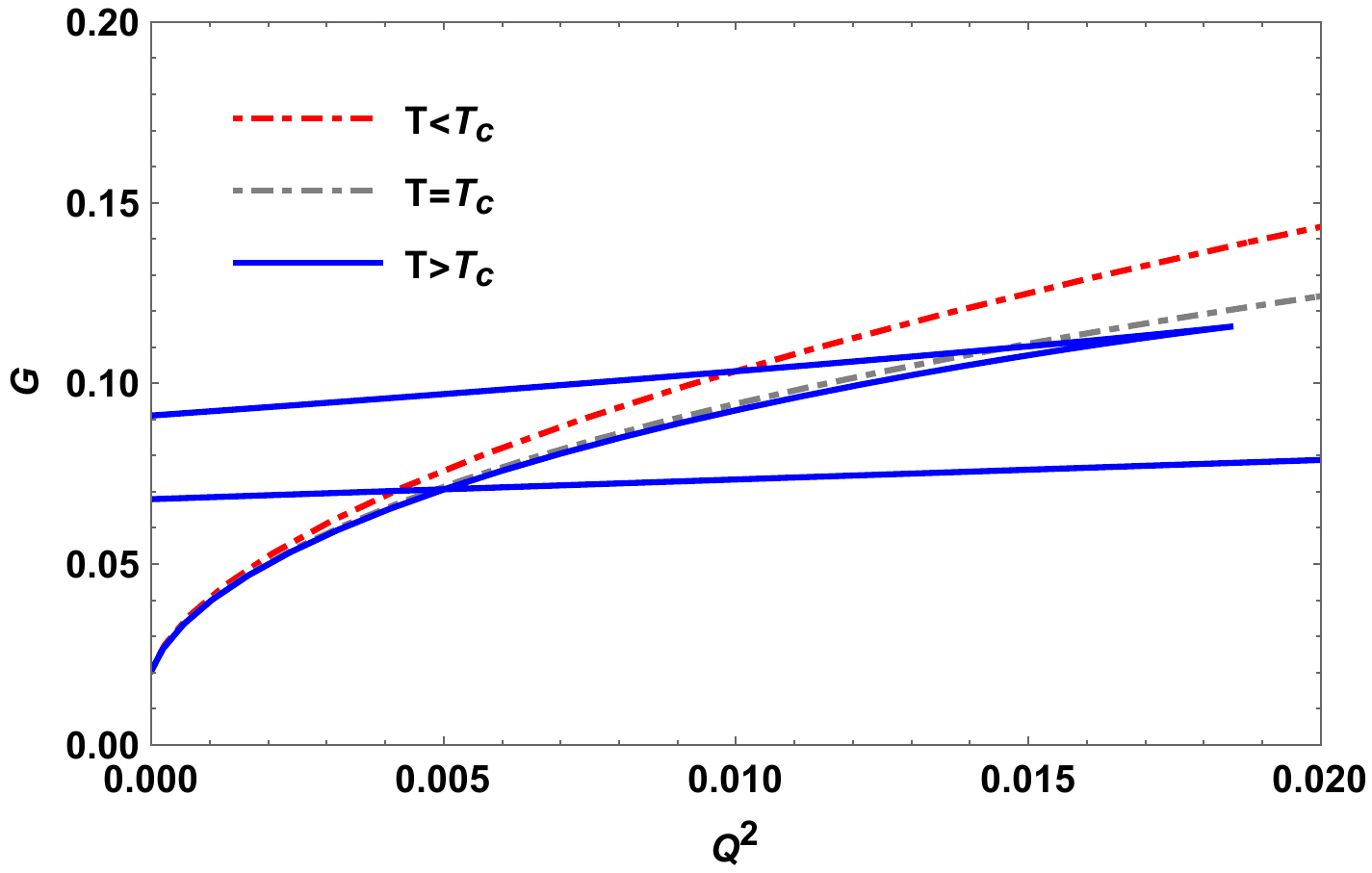} \vspace{-0.2cm}
        \includegraphics[scale = 0.32]{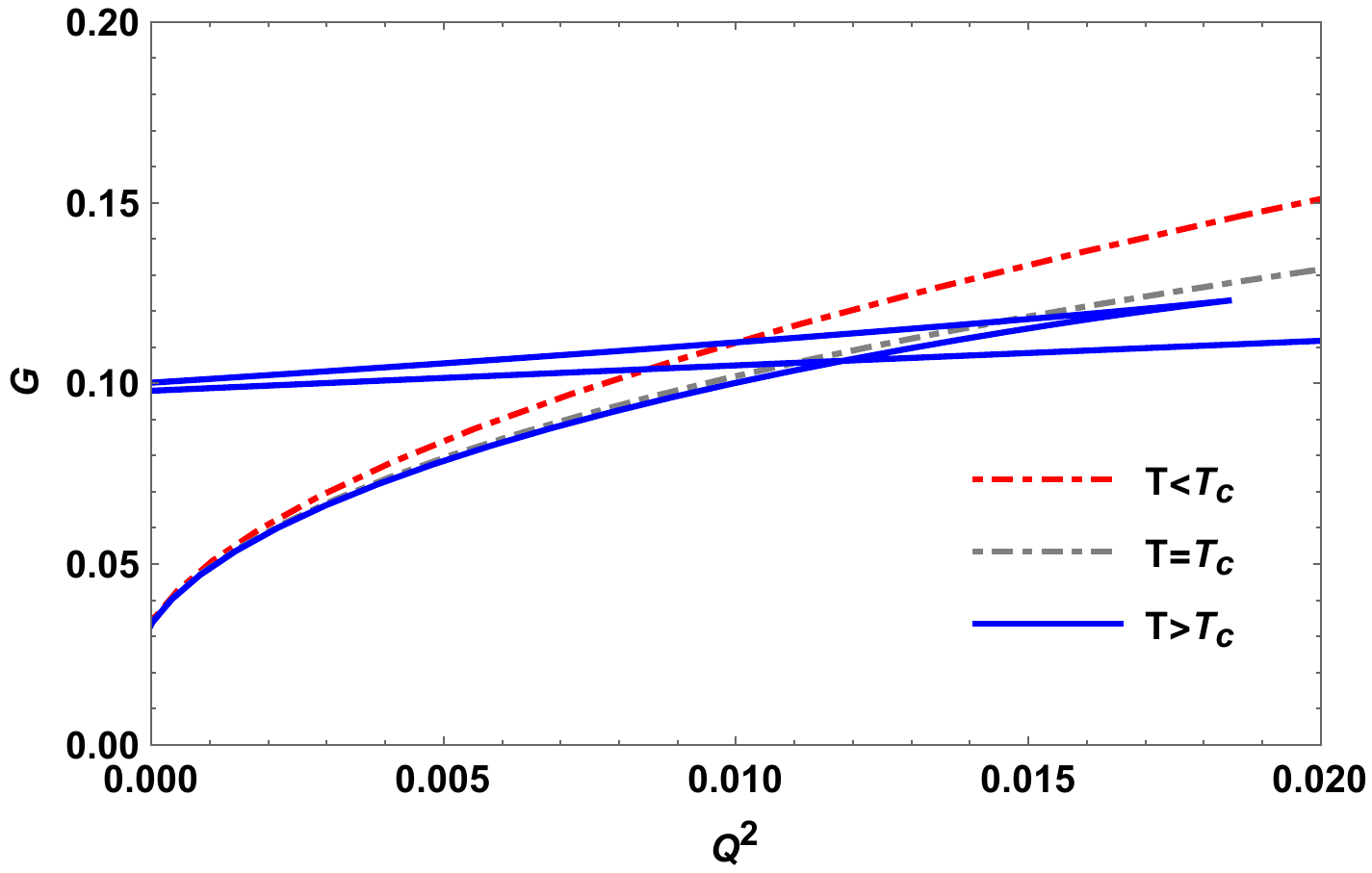} \vspace{-0.2cm}
		\caption{.$G-Q^{2}$ diagram Up panel:
 $(\beta=l=1,A=0,B=0.1,\gamma=0.01)$. Middle panel: $(\beta=2,l=1,A=0,B=0.1,\gamma=0.01)$. Down panel: $(A=B=\beta=0.1,l=1,\gamma=0.01$).}
		\label{fig:WKBL0}
	\end{figure}
Next, we proceed with the computation of critical exponents within this novel phase space methodology. The characteristics of thermodynamic functions in close proximity to the critical point are determined by the critical exponents, which is defined by
\begin{equation}
\begin{aligned}
&C_{\psi}=|t|^{-\tilde{\alpha}}, \quad \eta=|t|^\chi,\\
&\kappa_T=|t|^{-\tilde{\gamma}}, \quad\left|Q^2-Q_c^2\right|=\left|\psi-\psi_c\right|^\delta,
\end{aligned}
\end{equation}
here we use $\tilde{\alpha}$, $\tilde{\gamma}$ and $\chi$  to describe the critical exponent, the isothermal compression coefficient, and the order parameter, because these symbols are also described MCG parameters in this article. To determine the critical exponent, we establish the following dimensionless quantities
\begin{equation}\label{JJ2}
\psi_r=\frac{\psi}{\psi_c}, \quad Q_r^2=\frac{Q^2}{Q_c^2}, \quad T_r=\frac{T}{T_c}.
\end{equation}
Initially, we consider the temperature $T$ as a determinant of entropy
\begin{equation}
S=S(T, \psi)=\frac{\pi}{4\psi^2},
\end{equation}
we can find the temperature has no effect on entropy. If $\psi$ is fixed, the specific heat can be expressed as
\begin{equation}
C_{\psi}=\left.T \frac{\partial S}{\partial T}\right|_{\psi}=0.
\end{equation}
Based on this, $\tilde{\alpha}$ characterizes the behavior specific heat in close proximity to the critical point, we can obtain
\begin{equation}
 \tilde{\alpha}=0.
\end{equation}
To determine other critical exponents, we define the reduced variables as follows:
\begin{equation}
\psi_r=1+\epsilon, T_r=1+t, Q_r=1+\varrho.
\end{equation}
By taking into account the Eq. \eqref{JJ2},  the Eq. \eqref{JJ3} can be translated into a state equation in close proximity to critical points
\begin{equation}\label{JJ4}
\varrho=\Lambda_3 t-3 \Lambda_3 t \epsilon+\Lambda_5 \epsilon^3+0\left(t \epsilon^2+\epsilon^4\right),
\end{equation}
where

\begin{equation}
\begin{aligned}
& \Lambda_{0}=(1+A)(1+\beta), \\
& \Lambda_1=\frac{1}{16 \psi_c^4 Q_c^2}\left(\frac{1+A}{B}\right)^\alpha \frac{\alpha}{B}\left(8 \gamma \psi_c\right)^{\Lambda_0},\\
&\Lambda_2=\frac{1}{4 \psi_c^2 Q_c^2}, \quad \Lambda_3=\frac{\pi T_c}{2 \psi_c^3 Q_c^2}\\
&\Lambda_4=\frac{1}{16\psi_c^4 Q_c^2}\left[\frac{3}{1^2}-\left(\frac{1+A}{B}\right)^\alpha\right],\\
&\Lambda_5=\frac{\Lambda_1}{6}\left(3 \Lambda_{0}-6\right)\left(3 \Lambda_{0}-5\right)\left(3 \Lambda_{0}-5\right)\\
&\quad-4 \Lambda_2-10 \Lambda_3-20 \Lambda_4.
\end{aligned}
\end{equation}
By assuming $Q^{2}$ remains constant, and $t$ is a positive value, Based on Maxwell's equal area theory, we can obtain
\begin{equation}
\begin{aligned}
\varrho & =\Lambda_3 t-3 \Lambda_3 t \epsilon_l+\Lambda_5 \epsilon_l^3 \\
& =\Lambda_3 t-3 \Lambda_3 t \epsilon_s+\Lambda_5 \epsilon_s^3,
\end{aligned}
\end{equation}
\begin{equation}
\int_{\epsilon_{\mathrm{l}}}^{\epsilon_s} \epsilon\left(\Lambda_5-\Lambda_3 \epsilon^2\right) d \epsilon=0.
\end{equation}
Furthermore, the value of the critical exponent $\chi$ can be determined as follows:
\begin{equation}
\begin{aligned}
\left|\epsilon_s-\epsilon_l\right|=2 \epsilon_s & =\sqrt{\frac{3 \Lambda_3 t}{\Lambda_5}} \\
& \Rightarrow \chi=\frac{1}{2}.
\end{aligned}
\end{equation}
By substituting $t=0$, one can readily obtain the critical exponent $\delta$
\begin{equation}
\varrho=\Lambda_5 \epsilon_{\mathrm{k}}^3 \Rightarrow \delta=3.
\end{equation}
Finally, the critical exponents related to the behavior of the isothermal compression are calculated as follows
\begin{equation}
\kappa_T \propto \frac{\psi_c}{-3 \Lambda_3 Q_c^2 t}, \Rightarrow \tilde{\gamma}=1.
\end{equation}
Based on this, the findings indicate that the critical exponent of the system aligns with that obtained in the van der Waals fluid system \cite{cq9}.
\section{J–T Expansion }\label{sec4}
In this section,  we will study the J-T expansion of  charged AdS black hole accompanied by MCG. It is widely recognized that in AdS space, the mass of a black hole is considered as its enthalpy \cite{ch9,ch10}. During the expansion process, the temperature undergoes changes in response to pressure fluctuations while maintaining a constant enthalpy, the J-T coefficient is given by
\begin{equation}\label{JJ1}
\mu=\left(\frac{\partial T}{\partial P}\right)=\frac{1}{C_P}\left[T\left(\frac{\partial V}{\partial T}\right)-V\right],
\end{equation}
here, the positive and negative of J-T coefficient characterizes the cools $(\mu>0)$ or warms $(\mu<0)$ of gas adiabatic expansion \cite{ch12}. The heat capacity $C_{P}$  is defined by
\begin{figure}[htbp]
	\includegraphics[height=5.5cm,width=7.5cm]{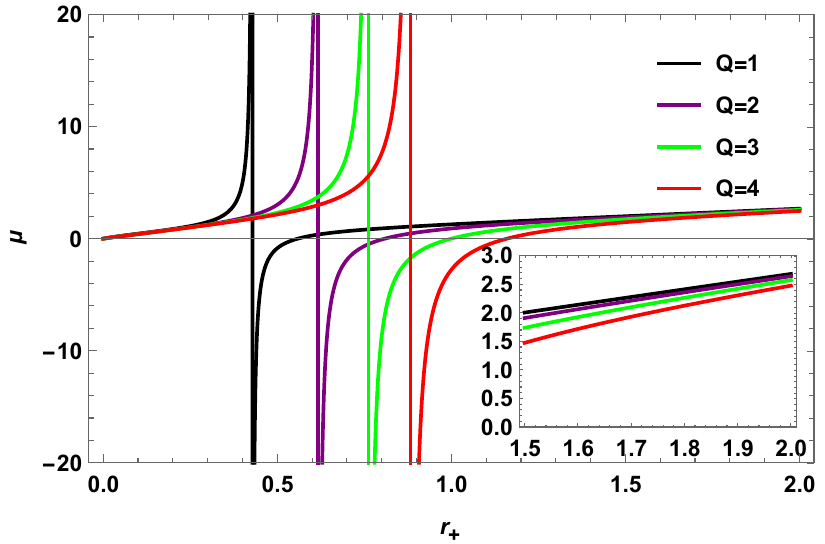}
	\caption{J-T coefficient $\mu$ of charged black holes for $P=A=1, B=\beta=\gamma=0.1$.}
\end{figure}
\begin{equation}
\begin{aligned}
C_p &= T\left(\frac{\partial S}{\partial T}\right)\\&=\frac{2\left(\frac{1+A}{B}\right)^\alpha B \pi r_{+}^6}{\left(\frac{1+A}{B}\right)^\alpha r_{+} \varrho_1+\varrho_2}+\frac{2 B \pi Q^2 r_{+}^2(1-\xi)^\alpha}{\left(\frac{1+A}{B}\right)^\alpha r_{+} \varrho_1+\varrho_2} \\
&-  \frac{2 B \pi r_{+}^4(1-\xi)^\alpha}{\left(\frac{1+A}{B}\right)^\alpha r_{+} \varrho_1+\varrho_2}-\frac{16 B P \pi^2 r_{+}^6(1-\xi)^\alpha}{\left(\frac{1+A}{B}\right)^\alpha r_{+} \varrho_1+\varrho_2}\\
&-\frac{2\left(\frac{1+A}{B}\right)^\alpha B \pi r_{+}^6 \xi}{\left(\frac{1+A}{B}\right)^\alpha r_{+} \varrho_1+\varrho_2}-\frac{2 B \pi Q^2 r_{+}^2(1-\xi)^\alpha \xi}{\left(\frac{1+A}{B}\right)^\alpha r_{+} \varrho_1+\varrho_2}\\
&+\frac{2 B \pi r^4_{+}(1-\xi)^\alpha \xi}{\left(\frac{1+A}{B}\right)^\alpha r_{+} \varrho_1+\varrho_2}+\frac{16 B P \pi^2 r^6_{+}(1-\xi)^\alpha \xi}{\left(\frac{1+A}{B}\right)^\alpha r_{+} \varrho_1+\varrho_2}
\end{aligned}
\end{equation}
with\\
$\varrho_1=\left(-3(1+A) \gamma\left(\frac{\gamma}{r_{+}^3}\right)^{A+\beta+A \beta}-B r_{+}^3(-1+\xi)\right),$\\
$\varrho_2=B\left(3 Q^2-r_{+}^2+8 P \pi r_{+}^4\right)(1-\xi)^\alpha(-1+\xi).$\\
\begin{figure}[htbp]
	\includegraphics[height=5.5cm,width=7.5cm]{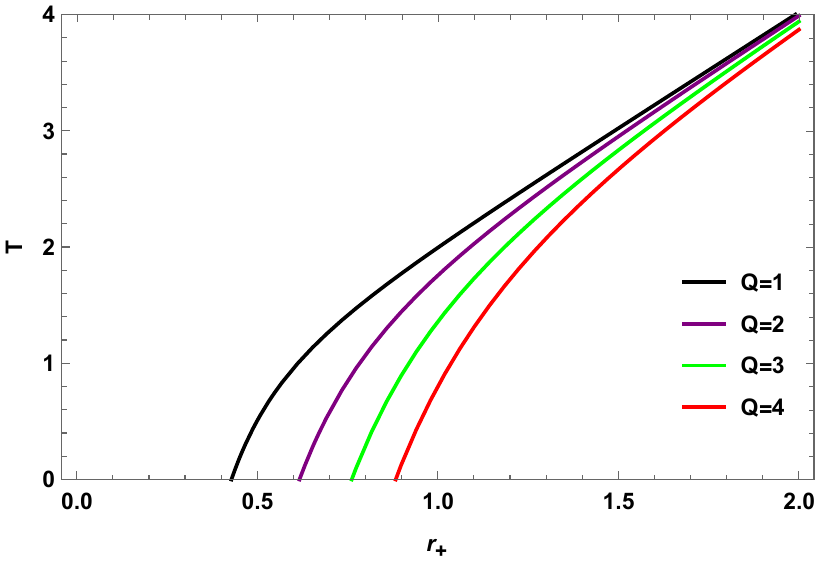}
	\caption{Hawking temperature of charged black holes for $P=A=1,B=\beta=\gamma=0.1$.}
\end{figure}
\begin{figure}[htbp]
	\includegraphics[height=5.5cm,width=7.5cm]{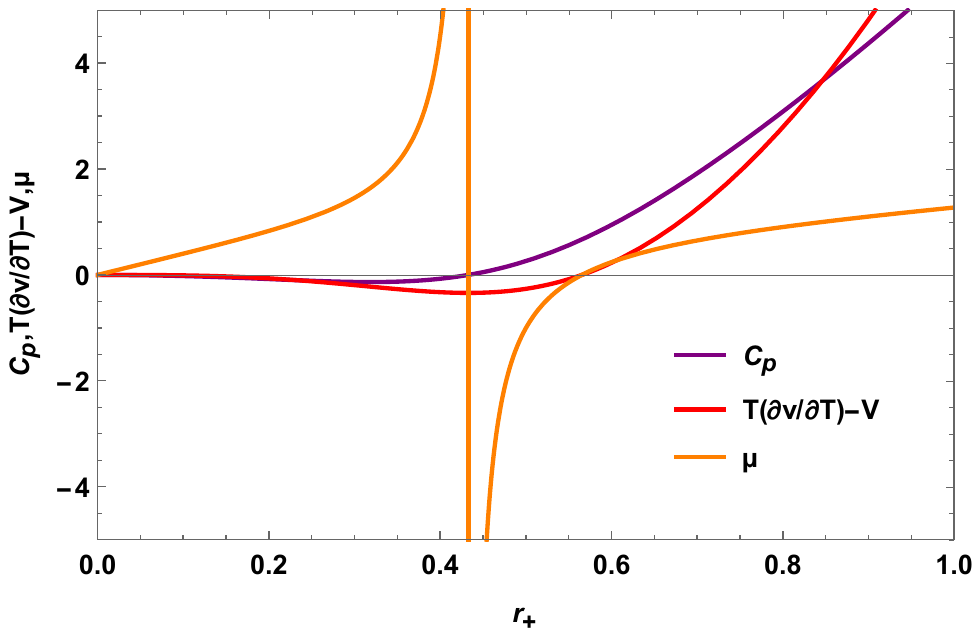}
	\caption{Graphical behavior of the J-T coefficient, heat capacity , and $\left[T\left(\frac{\partial V}{\partial T}\right)-V\right]$ of charged black holes for $Q=P=A=1, B=\beta=\gamma=0.1$.}
\end{figure}
We present the variation of $\mu$ and $T$ with the horizon radius $r_{+}$ in FIGs. 3-4, by setting $P=A=1,B=\beta=\gamma=0.1$, we can observe the constraints of the charge $Q$ on the J-T coefficient $\mu$ and temperature $T$, we can find that the divergence point ( vertical curve ) and the zero point move to the right as the charge increases. The zero point of the $T-r_{+}$ is the same as the $\mu-r_{+}$ divergence, the divergence points $(0.42846,0.61696,0.76147,0.88281)$ correspond to the charge values $1, 2, 3$, and $4$, respectively. In addition, we observe a decrease in the Hawking temperature as the charge $Q$ increases, and the thermodynamic characteristics of the black hole are greatly influenced by the charge.

Next, we will investigate the potential impact of negative heat capacity on J-T expansion. To this end, we present the changes in J-T coefficient and heat capacity as a function of horizon radius, as depicted in FIG.5.  The results show that the heat capacity has only one root ($Q=P=A=1,B=\beta=\gamma=0.1$, $r_{0}=0.42846)$. If the horizon radius $r_{+}$ is smaller than $r_{0}(C_{p}=0)$, it can be observed that the heat capacity exhibits negativity and the J-T coefficient demonstrates positivity, suggesting the presence of a cooling process. In addition, we can observe that if the heat capacity and $\left[T\left(\frac{\partial V}{\partial T}\right)-V\right]$  are negative, $\mu$ will be positive, which can perfectly interpret the Eq.\eqref{JJ1}. If the J-T coefficient is reduced to zero, one can determine the values of inversion temperature and inversion pressure, which can be represented as follows:
\begin{equation}
\begin{aligned}
T_i& =V \frac{\partial T}{\partial V} \\&= \frac{Q^2}{4 \pi r_{+}^3}-\frac{1}{12 \pi r_{+}}-\frac{\left(\frac{1+A}{B}\right)^\alpha(1-\xi)^{-\alpha} r_{+}}{12 \pi} \\
& +\frac{\left(\frac{1+A}{B}\right)^\alpha \gamma(1-\xi)^{-1-\alpha}\left(\frac{\gamma}{r_{+}^3}\right)^{A+\beta+A \beta}}{4 B \pi r_{+}^2}+\frac{2 P r_{+}}{3}\\
&+\frac{A\left(\frac{1+A}{B}\right)^\alpha \gamma(1-\xi)^{-1-\alpha}\left(\frac{\gamma}{r_{+}^3}\right)^{A+\beta+A \beta}}{4 B \pi r_{+}^2},
\end{aligned}
\end{equation}
\begin{equation}
\begin{aligned}
P_i & =\frac{\left(\frac{1+A}{B}\right)^\alpha(1-\xi)^{-1-\alpha}}{8 \pi}-\frac{5\left(\frac{1+A}{B}\right)^\alpha(1-\xi)^{-1-\alpha} \xi}{16 \pi} \\
& +\frac{3 Q^2}{8 \pi r_+^4}-\frac{3 A\left(\frac{1+A}{B}\right)^\alpha(1-\xi)^{-1-\alpha} \xi}{16 \pi}-\frac{1}{4 \pi r_+^2}.
\end{aligned}
\end{equation}

\begin{figure}
		\centering
		\includegraphics[scale = 0.58]{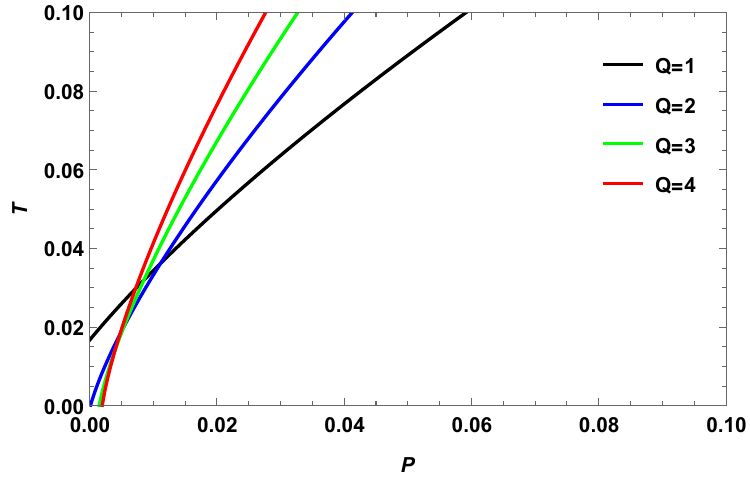} \hspace{-0.2cm}
		\includegraphics[scale = 0.58]{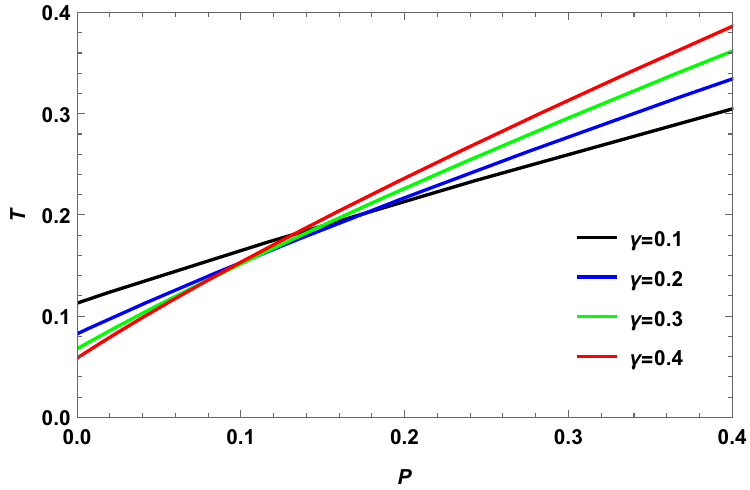} \vspace{-0.2cm}
        \includegraphics[scale = 0.58]{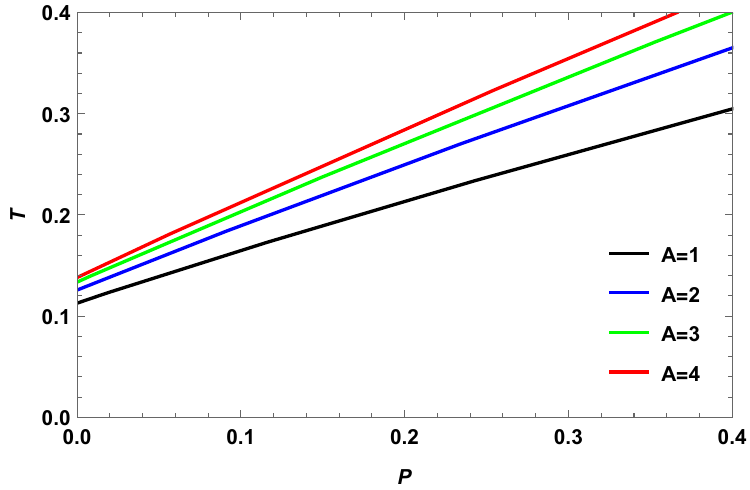} \vspace{-0.2cm}
        \includegraphics[scale = 0.58]{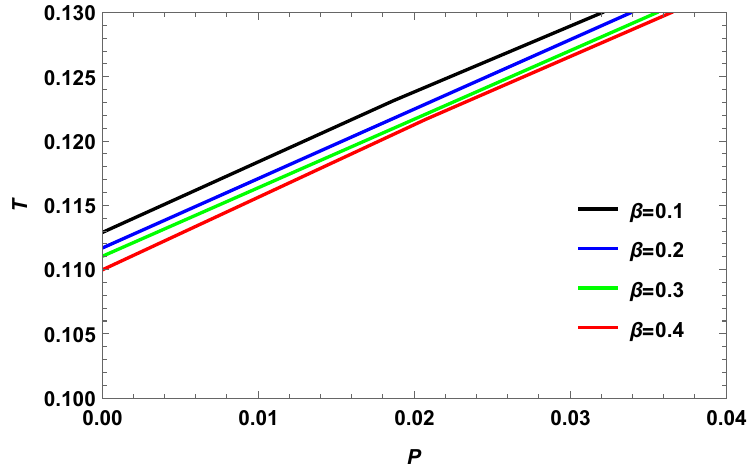} \vspace{-0.2cm}
		\caption{The inversion curve for different parameter spaces.  From top to bottom, we set ($A=1,B=\beta=\gamma=0.1$), ($A=1,B=\beta=Q=0.1$), ($\beta=\gamma=B=Q=0.1$) and ($A=1, \gamma=B=Q=0.1$).}
		\label{fig:WKBL0}
	\end{figure}
The effects of different values of charge $Q$ and parameter $\gamma$, $A$ and $\beta$ on the inversion curves can be observed in FIG.6. From top to bottom , we set ($A=1,B=\beta=\gamma=0.1$), ($A=1,B=\beta=Q=0.1$), ($\beta=\gamma=B=Q=0.1$) and ($A=1, \gamma=B=Q=0.1$), respectively. From the previous two images, we find that charge $Q$ and parameter $\gamma$ have similar effects on the inversion curve, that is, the increase in charge $Q$ and parameter $\gamma$ leads to a decrease in the inversion curve in low pressure, while the opposite behavior exists under high pressure. From the latter two images, it can be seen that parameters  $A$ and $\beta$ have opposite effects on the inversion curve. Interestingly, the presence of mixed dark matter and dark energy yields a similar influence on charged black holes as most multidimensional black hole systems \cite{ch13}. The presence of a minimum inversion temperature, at which $P_{i}$ becomes zero, is evident from FIG.6.

Next, the focus of our study is to investigate the ratio problem associated with the minimum inversion temperature. By ensuring that $P_{i}$ equals zero, we can derive the $r_{min}$ and $T_{min}$ as follows:
\begin{equation}
\begin{aligned}
& \left(\frac{1+A}{B}\right)^\alpha(1-\xi)^{-1-\alpha}(2-(5+3 A) \xi) \\
& +\frac{6 Q^2}{r_{min}^4}-\frac{4}{r_{min}^2}=0,
\end{aligned}
\end{equation}
\begin{equation}
\begin{aligned}
T_{min} & =\frac{Q^2}{2 \pi r_{min}^3}-\frac{3\left(\frac{1+A}{B}\right)^\alpha(1-\xi)^{-1-\alpha} \xi r_{min}}{8 \pi} \\
& -\frac{3 A\left(\frac{1+A}{B}\right)^\alpha(1-\xi)^{-1-\alpha} \xi r_{min}}{8 \pi}-\frac{1}{4 \pi r_{min}}.
\end{aligned}
\end{equation}
In addition, the expression for the critical temperature can be calculated by $\left(\partial_{r_+} P\right)_{T}=0$, $\left(\partial_{r_+,r_+} P\right)_{T}=0$
\begin{equation}
\begin{aligned}
T_{\mathrm{c}}&=  \frac{1}{4 \pi r_c^2}+\frac{1}{2 \pi r_c}-\frac{Q^2}{\pi r_c^3}-\frac{32^{-2+\frac{\beta}{1+\beta}} \beta \gamma^2}{\pi r_c^5} \\
&+ \frac{\beta\left(1+\left(\frac{\gamma}{r_c^3}\right)^{2(1+\beta)}\right)^{-1+\frac{1}{1+\beta}}}{4 \pi r_c^2}.
\end{aligned}
\end{equation}
In FIG.7, we can observe an initial increase followed by a subsequent decrease in the ratio $\frac{T_{min}}{T_{c}}$ as the charge increases. For smaller $Q$, the ratio increases with the increase of parameters $\beta$ and $\gamma$. When $Q$ is increased, the ratio decreases with the increase of parameters $\beta$ and $\gamma$. For larger $Q$, the increase of $\beta$ parameter will decrease the zero point of the ratio. However, the increase of $\gamma$ has no effect on the zero point of the ratio.
\begin{figure}
		\centering
		\includegraphics[scale = 0.6]{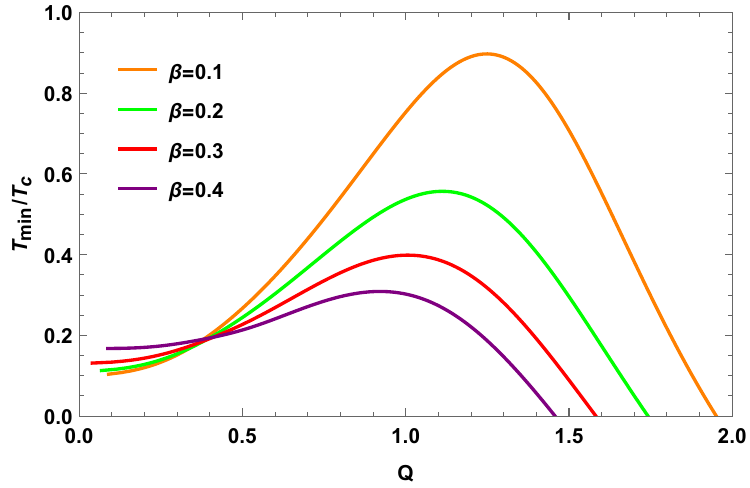} \hspace{-0.2cm}
		\includegraphics[scale = 0.6]{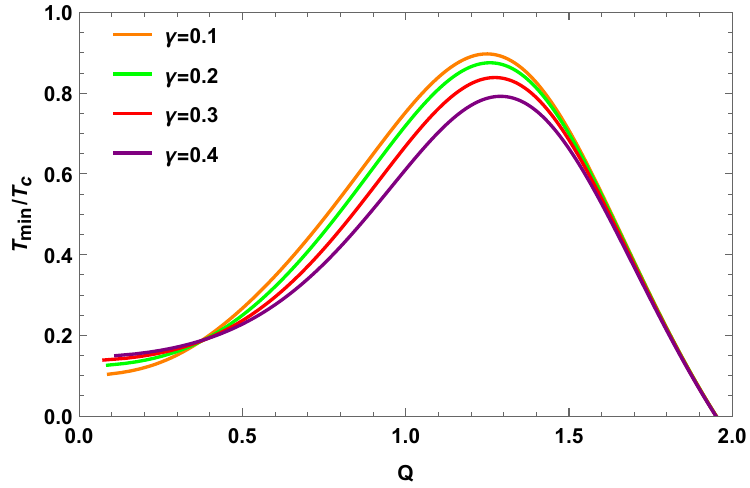} \vspace{-0.2cm}
        \caption{Effect of the parameters $\beta$ and $\gamma$ on the ratio $\frac{T_{min}}{T_{c}}$. Up panel: ($A=1,B=\gamma=0.1$). Down panel: ($A=1,B=\beta=0.1$). }
		\label{fig:WKBL0}
	\end{figure}

Next, we will conduct an analytical examination of the ratio in certain specific scenarios, such as, for $\beta = 0$, $\gamma = 0$, and $B\rightarrow0$, $r_{min}$ has two roots as follows:
\begin{equation}
r_{min}=-\sqrt{\frac{3}{2}} Q, \quad\quad r_{min}=\sqrt{\frac{3}{2}} Q,
\end{equation}
we consider positive roots, the ratio is expressed as
\begin{equation}
\frac{T_{min }}{T_c}=\frac{Q}{\sqrt{6}\left(A+\sqrt{\frac{2}{3}} Q\right)}.
\end{equation}
If the parameter $A = 0$, the ratio $\frac{T_{min }}{T_c}=\frac{1}{2}$, it is completely consistent with the results of the charged AdS black hole \cite{ch11}.
\begin{figure}
		\centering
		\includegraphics[scale = 0.42]{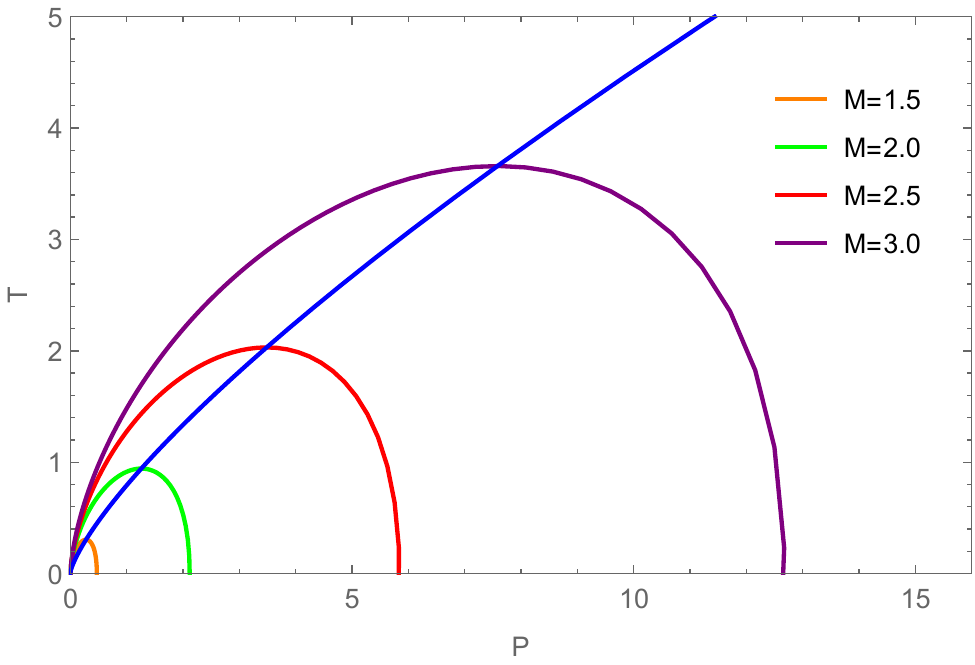} \hspace{-0.2cm}
		\includegraphics[scale = 0.42]{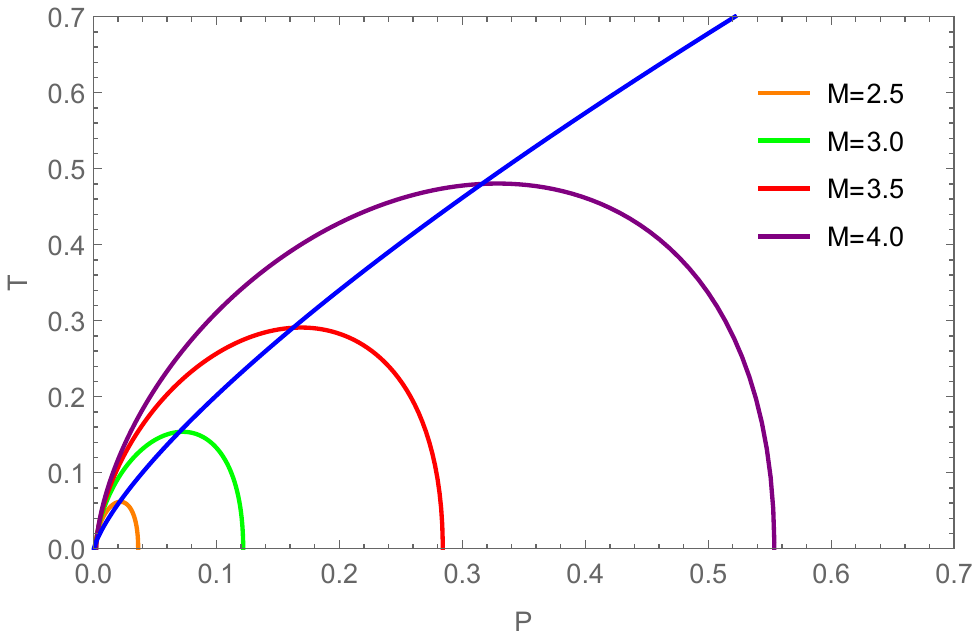} \vspace{-0.2cm}
        \includegraphics[scale = 0.42]{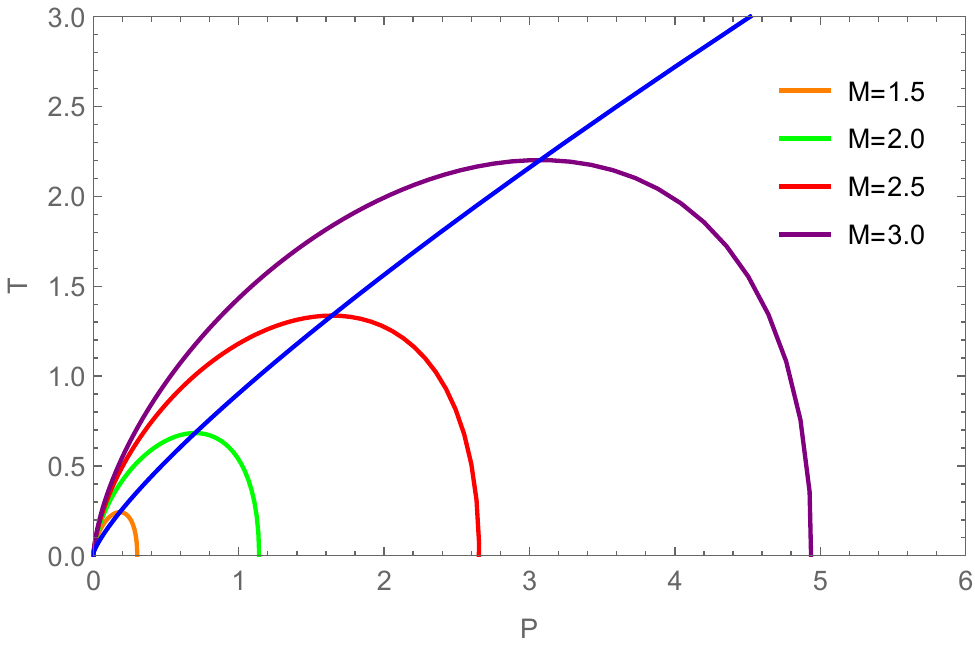} \vspace{-0.2cm}
         \includegraphics[scale = 0.42]{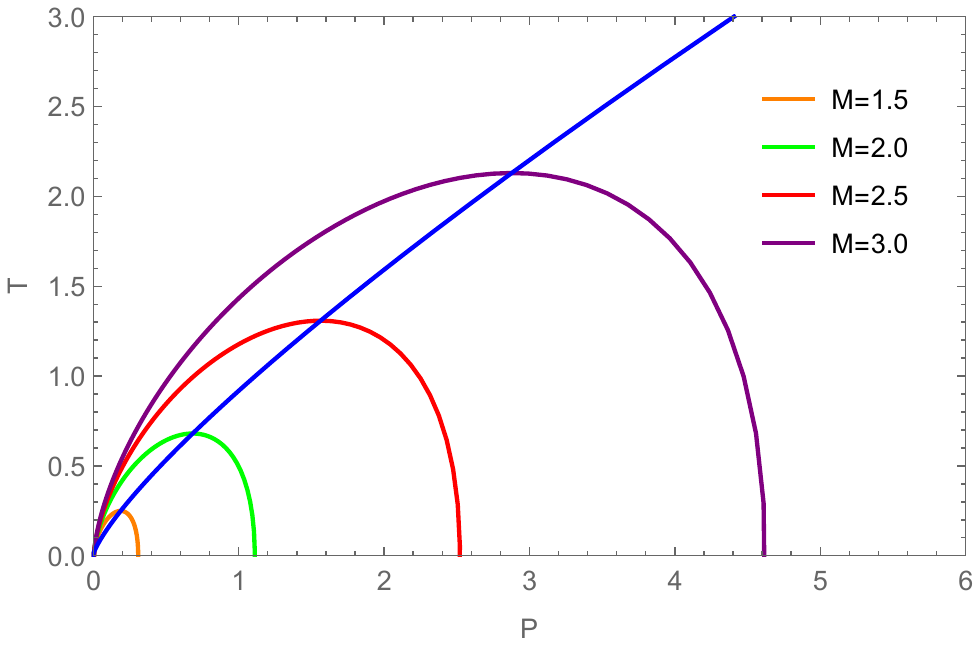} \vspace{-0.2cm}
		\caption{Isenthalpic curves (Non-blue) and inversion curves (blue). From top to bottom , we set($\beta=\gamma=B=0.1,A=Q=1$), ($\beta=\gamma=B=0.1,A=1, Q=2$), ($\beta=0.1, B=0.1,\gamma=0.5, A=Q=1$) and ($\beta=0.5, B=0.1,\gamma=0.5, A=Q=1$).}
		\label{fig:WKBL0}
	\end{figure}

Next, we will further examine the isenthalpic curve. From FIG.8, it can be observed that the inversion curve (blue curve) coincides with the maximum value of the isenthalpy curve (Non-blue curve). We can observe a distinct region where the slope of the constant enthalpy curve exceeds that of the inverse temperature curve, suggesting a cooling phenomenon takes place in this particular area. The sign of slope on the isenthalpic curve changes under inverse temperature curves, suggesting warming signs in certain regions. Determine the demarcation line between regions of heating and cooling for black holes using the inversion curve. We observe an upward trend in the inversion point as mass increases, while a significant decline is observed with increasing charge $Q$. The influence of MCG parameters on the inversion point is relatively small but cannot be ignored.
	\section{Conclusion}
In summary, the main focus of our research lies in the investigation of charged AdS black hole surrounded by an exotic fluid, with MCG as our primary subject. The phase transition and critical behavior of the system are thoroughly studied. Additionally, we investigate the influence of MCG parameters and the charge on the J-T expansion in the extended phase space. Our main findings are as follows:\\

$(i)$ The presence of an anomalous negative $Q^{2}$ component can be observed within this region on the $Q^{2}-\psi$ diagram when the Hawking temperature falls below the critical temperature. The black hole remains stable as long as the Hawking temperature does not exceed the critical temperature. Whereas, when the Hawking temperature exceeds the critical temperature, A characteristic pattern of swallowtail behavior is observed, which means the emergence of the first-order phase transition. Note that the critical exponent of the system is found to be in complete agreement with that observed in the van der Waals fluid system.\\

$(ii)$  The displacement of the divergence point (vertical curve) and the zero point shifts to the right with increasing charge. The $\mu-r_{+}$ divergence coincides with the $T-r_{+}$ zero point, the heat capacity is negative  indicates the occurrence of a cooling process.  At low pressure, an increase in charge $Q$ and parameter $\gamma$ leads to a decrease in the inversion temperature. Conversely, at high pressure, an increase in charge $Q$ and parameter $\gamma$ results in an increase in the inversion temperature.\\

$(iii)$ The parameters $\beta$ and $\gamma$ has a non-negligible effect on the ratio $\frac{T_{min}}{T_{c}}$. For larger $Q$, increasing the $\beta$ parameter will result in a decrease of the zero point of the ratio. However, altering the $\gamma$ has no impact on the zero point of the ratio.
We also observe that the parameters $Q$ and $M$ have opposite effects on the inversion poin. The change of MCG parameters  have a relatively small effect on the inversion point.	
	\vspace{5mm}
	\begin{acknowledgments}
		We would like to thank the anonymous referees for their valuable comments on improving our paper. This work is supported by the Doctoral Foundation of Zunyi Normal University of China (BS [2022] 07, QJJ-[2022]-314), and the National Natural Science Foundation of China (Grant Nos. 12265007 and 12264061). The research partially was supported by the Long-Term Conceptual Development of a University of Hradec Kr\'alov\'e for 2023, issued by the Ministry of Education, Youth, and Sports of the Czech Republic.
	\end{acknowledgments}

\end{document}